\documentclass[%
 aip,
 sd,%
 amsmath,amssymb,
 reprint,%
]{revtex4-1}

\usepackage{graphicx}
\usepackage{dcolumn}
\usepackage{bm}
\usepackage{color} 

\begin{document}


\title{
 Particle detection through the quantum counter concept in YAG:Er$^{3+}$
}

\author{A. F. Borghesani}
\affiliation{ 
CNISM Unit, Dip. di Fisica e Astronomia and INFN, Via F. Marzolo 8, I-35131 Padova, Italy
}
\author{C. Braggio}
 \email{Caterina.Braggio@unipd.it}
\affiliation{ 
Dip. di Fisica e Astronomia and INFN, Via F. Marzolo 8, I-35131 Padova, Italy
}
\author{G. Carugno}
\affiliation{ 
Dip. di Fisica e Astronomia and INFN, Via F. Marzolo 8, I-35131 Padova, Italy
}

\author{F. Chiossi}
\affiliation{ 
Dip. di Fisica e Astronomia and INFN, Via F. Marzolo 8, I-35131 Padova, Italy
}

\author{A. Di Lieto}
\affiliation{ 
Dip. di Fisica and INFN, Largo Bruno Pontecorvo, 3, I-56127 Pisa, Italy
}

\author{M. Guarise}
\affiliation{ 
Dip. di Fisica e Astronomia and INFN, Via F. Marzolo 8, I-35131 Padova, Italy
}

\author{G. Ruoso}
\affiliation{ 
INFN, Laboratori Nazionali di Legnaro, Viale dell'Universit\`a 2, I-35020 Legnaro, Italy
}

\author{M. Tonelli}
\affiliation{ 
Dip. di Fisica and INFN, Largo Bruno Pontecorvo, 3, I-56127 Pisa, Italy
}


\begin{abstract}

We report about a novel scheme for particle detection based on the infrared quantum counter concept. Its operation consists of a two-step excitation process of a four level system, that can be realized in rare earth-doped crystals when a cw pump laser is tuned to the transition from the second to the fourth level. The incident particle raises the atoms of the active material into a low lying, metastable energy state, triggering the absorption of the pump laser to a higher level. 
Following a rapid non-radiative decay to a fluorescent level, an optical signal is observed with a conventional detectors.
In order to demonstrate the feasibility of such a scheme, we have investigated the emission from the fluorescent level $^4$S$_{3/2}$ (540\,nm band) in an Er$^{3+}$-doped YAG crystal pumped by a tunable titanium sapphire laser when it is irradiated with 
60 keV electrons delivered by an electron gun. We have 
obtained a clear signature this excitation increases the $^{4}I_{13/2}$ metastable level population that can efficiently be  exploited to generate a detectable optical signal.

 
\end{abstract}

\maketitle

There is a significant interest in the development of devices for the detection of low rate, low energy deposition events  both for dark matter searches \cite{Sikivie:2014fk} and the study of neutrino interactions in condensed matter \cite{Dodd:1991uq,Freedman:1974fk}. A very low value of energy threshold ($\sim 0.5$ keV) has been reported in semiconductor detectors \cite{Aalseth:2011fk}, or in bolometers, where a few tens of eV events could be detected \cite{enss:2005}. 
Unfortunately, their active mass, a crucial parameter in rare events searches, are also tiniest: in the most sensitive detector only a few grams of material act as a target.

\begin{figure}[h!]
\begin{center}
\includegraphics[width=2.7in]{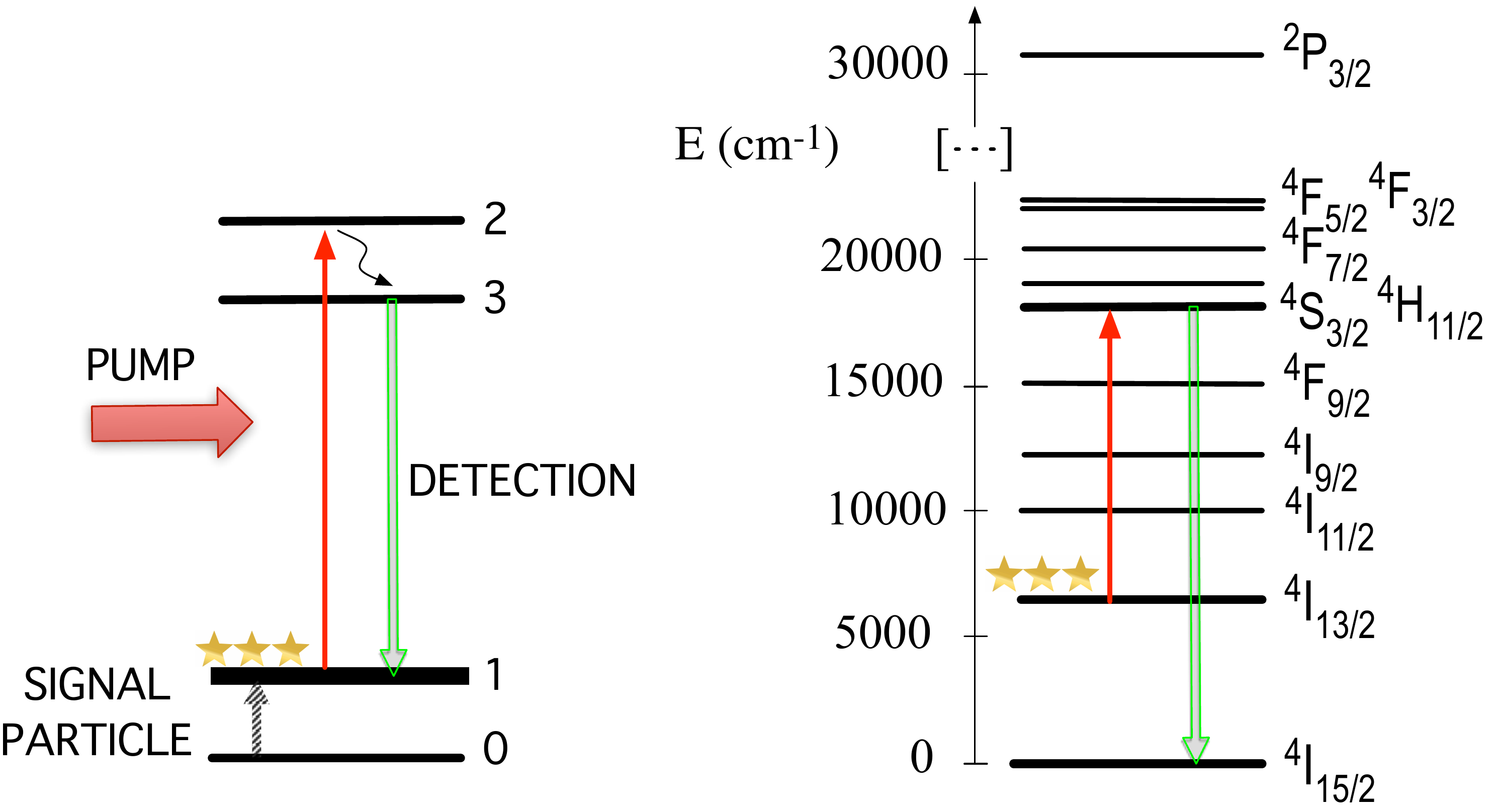}
\caption{\emph{(left)} Ideal four-level active material. \emph{(right)} Energy level scheme in YAG:Er \cite{gruber:1993}.}
\label{3liv}
\end{center}
\end{figure}

In the present work an all-optical detection scheme is proposed  to efficiently convert the incident particle energy into detectable photons. It is based on the infrared quantum counter concept (IRQC), proposed by N. Bloembergen \cite{blo:1959} as a way to extend photon detection to the  $1-100\,\mu$m wavelength range. The incident infrared photon is upconverted in a material that exhibits a four energy level system with $E_2>E_3>E_1>E_0$, as those determined in wide bandgap materials doped with trivalent rare-earth (RE) ions \cite{kaminski},  and kept under the action of a pump laser source resonant with transition $1\rightarrow2$. In analogy with the infrared quantum counter, detection of the particle is then accomplished through the fluorescence photons emitted in the transition $3\rightarrow1$ as shown in Fig.\,\ref{3liv}. 

In contrast to narrow-band, selective detection of infrared photons, a particle that interacts
 in an optical material gives rise to several phenomena, including energy transfer  processes from the host to the RE ion, which can be viewed as wideband excitation for the present purposes. 
  The complex chain of events whereby a particle loses its energy in a material has been systematically investigated for the development of state-of-the-art scintillators\cite{Rodnyi,Weber200235,Drozdowski:2008,moses:1998}. Attention has been paid to the transitions in the visible, UV and near infrared in those studies. Here we want to focus on the fraction of the particle energy that is translated in the excitation of the low energy metastable level 1 indicated in Fig.\,\ref{3liv}, that can take place both through the decay of higher levels and directly from the ground state. This is motivated by the assumption that the particle energy loss is a process in which it is much more probable to increase the population of low energy atomic levels than highest ones. 
  Such reasoning is supported by three points: 
  \begin{enumerate}
  \item the inelastic scattering of free electrons off bounded electrons is a major process in particle energy loss and is described by the Bethe-Bloch formula that privileges \emph{low energy transfer} events \cite{knoll};
  \item  another dominant process is the thermalization of the secondary electrons produced in the interaction that takes place through optical \emph{phonon scattering};
  \item  \emph{infrared scintillation} in the range $600-900\,$nm has been reported \cite{moses:1998} with an intensity in excess of $10^5$ photons per MeV. A light yield of $(79\pm8)\times10^{3}$ has also been observed in YAG:(10\%)Yb$^{3+}$, whose emission peaked around $\lambda=1.03\,\mu$m \cite{Antonini:2002fk}. 
   \end{enumerate}
     
  In addition, if we choose a material with a low energy level ($E_1\sim10-100$\,meV, corresponding to the wavelength range $\sim 100-10 \,\mu$m) characterized by a long lifetime, such metastable level acts as a resevoir in which the energy of the particle can be stored and converted in fluorescent photons by the pump laser. 
 In view of the previous hypotheses, and provided the efficiency of the upconversion process is high \cite{krupke:1965, wright:1973}, the proposed scheme has the potential to generate a greater number of information carriers (photons) for a given energy release.
 
 The present work is organized as follows. We first investigate the response of a low concentrated YAG:(0.5\%)Er$^{3+}$ crystal to an electron gun excitation, which represent our particle signal in Fig.\,\ref{3liv}. 
These preliminary measurements include both the acquisition of cathodoluminesce (CL) spectra and the study of the  $^{4}I_{13/2}$ metastable lifetime. 
We also estimate the efficiency of the laser-pumped YAG:Er crystal in the counting of photons delivered by a diode laser ($\lambda \approx 960$\,nm). 
 The core of the work is the study of the fluorescence stemming from the de-excitation of the $^{4}S_{3/2}$ level when the particle-excited crystal is continuously pumped by a laser resonant with the transition  $^{4}I_{13/2}\rightarrow ^{4}S_{3/2}$.


 
\begin{figure}[h!]
\begin{center}
\includegraphics[width=3.1in]{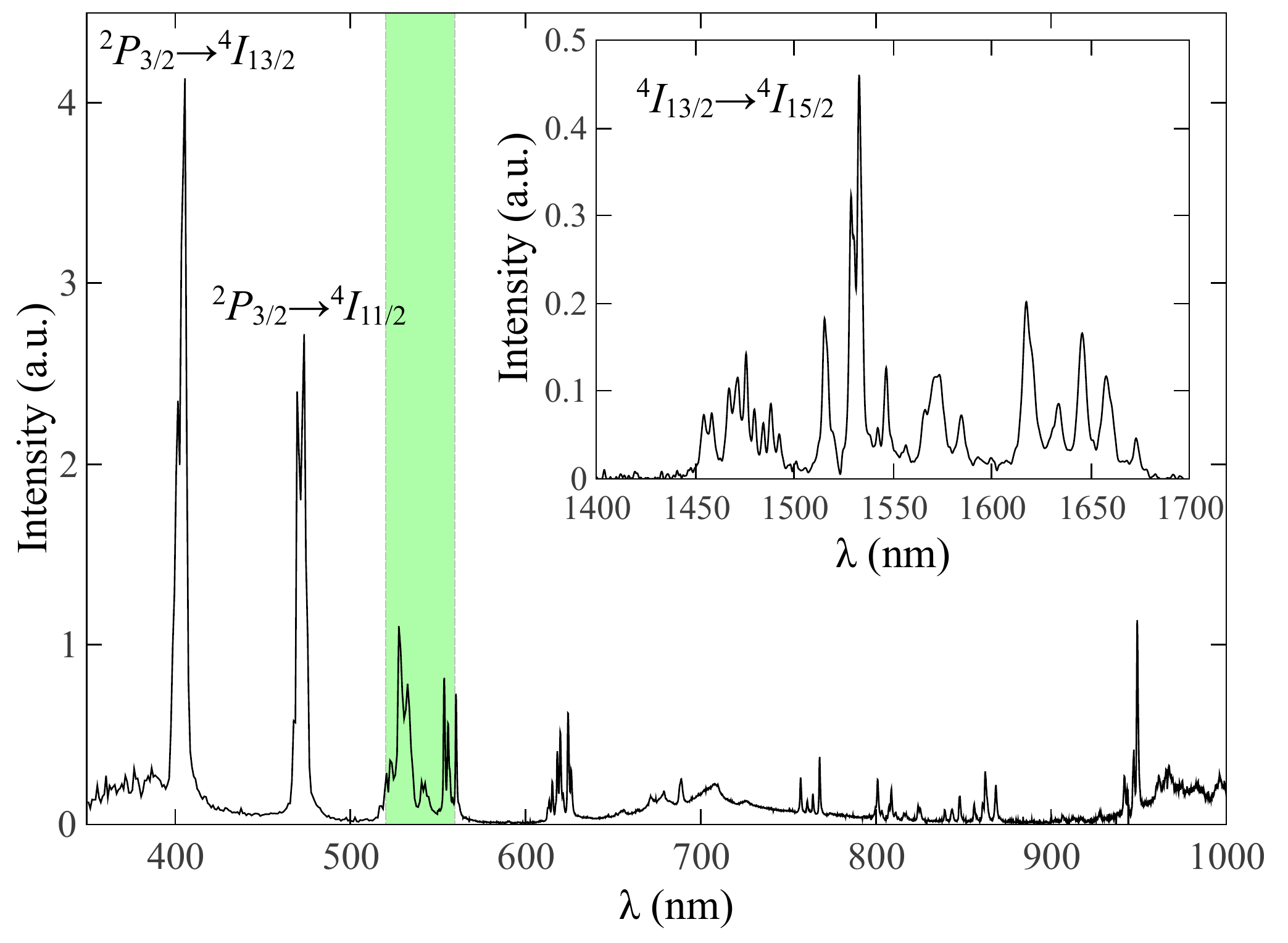}
\caption{CL spectra of YAG:Er. The results of Fourier transform interferometer (FT-IR) measurements are displayed in the inset.}
\label{virsp}
\end{center}
\end{figure}

During the CL measurements the YAG:Er crystal, a cylinder of 5\,mm diameter and height 3\,mm, is mounted at the far end of the electron gun vacuum chamber that is enclosed inside a Pb-shielded cabinet \cite{borghe:2011}. The main transitions involved in the excitation process can be identified in the spectra shown in Fig.\,\ref{virsp}. The visible spectrum is obtained with a CCD spectrometer (Ocean Optics mod.\,Red Tide 650), whereas the  
infrared portion of the spectrum, displayed in the inset, is studied with a Fourier transform interferometer (Bruker Equinox 55). The interferometer is equipped with an InGaAs photodiode, sensitive in the range $0.8-1.7\,\mu$m. 
The two sharp emission lines observed in the visible are doublets peaking at $401.5, 405.1$\,nm and at $470.9, 474.0$\,nm. They  are attributed to $4f-4f$ transitions of the Er$^{3+}$ ions in the YAG host, respectively to intermanifold transitions $^{2}P_{3/2}\rightarrow ^{4}I_{13/2}$ and $^{2}P_{3/2}\rightarrow ^{4}I_{11/2}$. 
These transitions indicate that the metastable level $^{4}I_{13/2}$ population is increased also by the decay of higher levels excited by electron impact.

As far as the 540\,nm band luminescence is concerned, CL spectra in YAG:Er single crystalline films have very recently been reported that show a larger emission\cite{Zorenko:2014fk} than in the present case. This might indicate that a portion of our $540$\,nm fluorescence band is absorbed in the 3\,mm-thick crystal, where the electrons interaction region is limited to a few hundred $\mu$m \cite{nist} from the surface while light detection is accomplished on the opposite crystal face. 

\begin{figure}[h!]
\begin{center}
\includegraphics[width=2.9in]{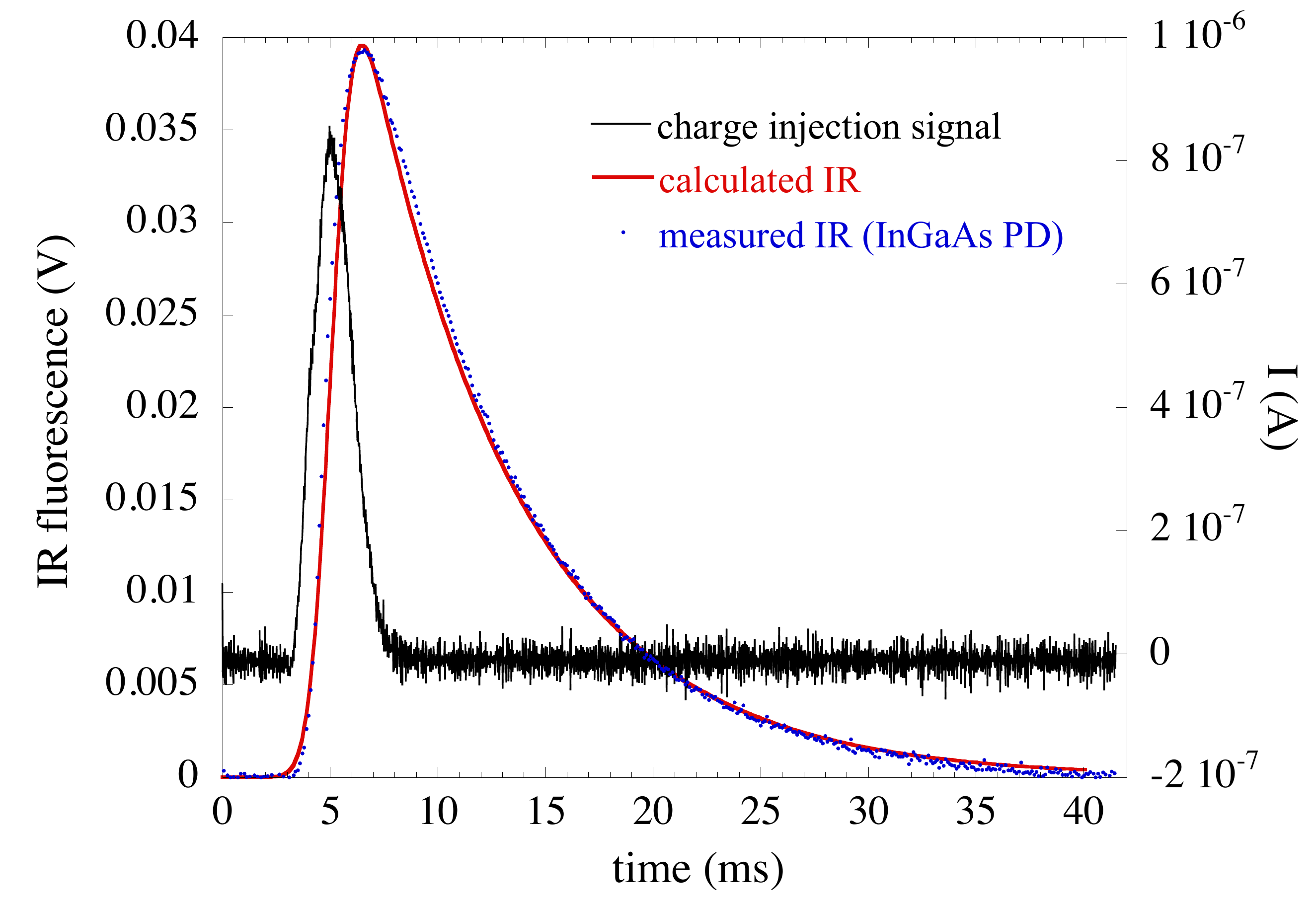}
\caption{Electron current waveform (right scale) and the 1.53\,$\mu$m fluorescence signal (left scale) detected by an InGaAs photodiode with input bandpass filter centered at $\lambda=1.490\pm0.25\,\mu$m. }
\label{life}
\end{center}
\end{figure}

In the inset we show the infrared spectrum originating from the decay of the  $^{4}I_{13/2}$ metastable level. This manifold around 1.53\,$\mu$m is widely used for IR lasing\cite{eich:2008}. The lifetime of this level is several milliseconds long and weakly depends on the crystal preparation and excitation source. Therefore we have recorded, as shown in Fig.\,\ref{life},  
the time evolution of the fluorescence emitted by the crystal excited with a $2.5\,$ms-duration electron pulse. 
The crystal response has been numerically computed by solving a first order kinetic equation for the population of the level excited by the current pulse shown in the figure. We obtain a lifetime value $\tau=7.19\pm0.02$ \,ms, in fair agreement with literature data \cite{Payne:1992,Ter-Gabrielyan:2009,toma:2006} obtained with optical excitation. 

\begin{figure}[h!]
\begin{center}
\includegraphics[width=3.0in]{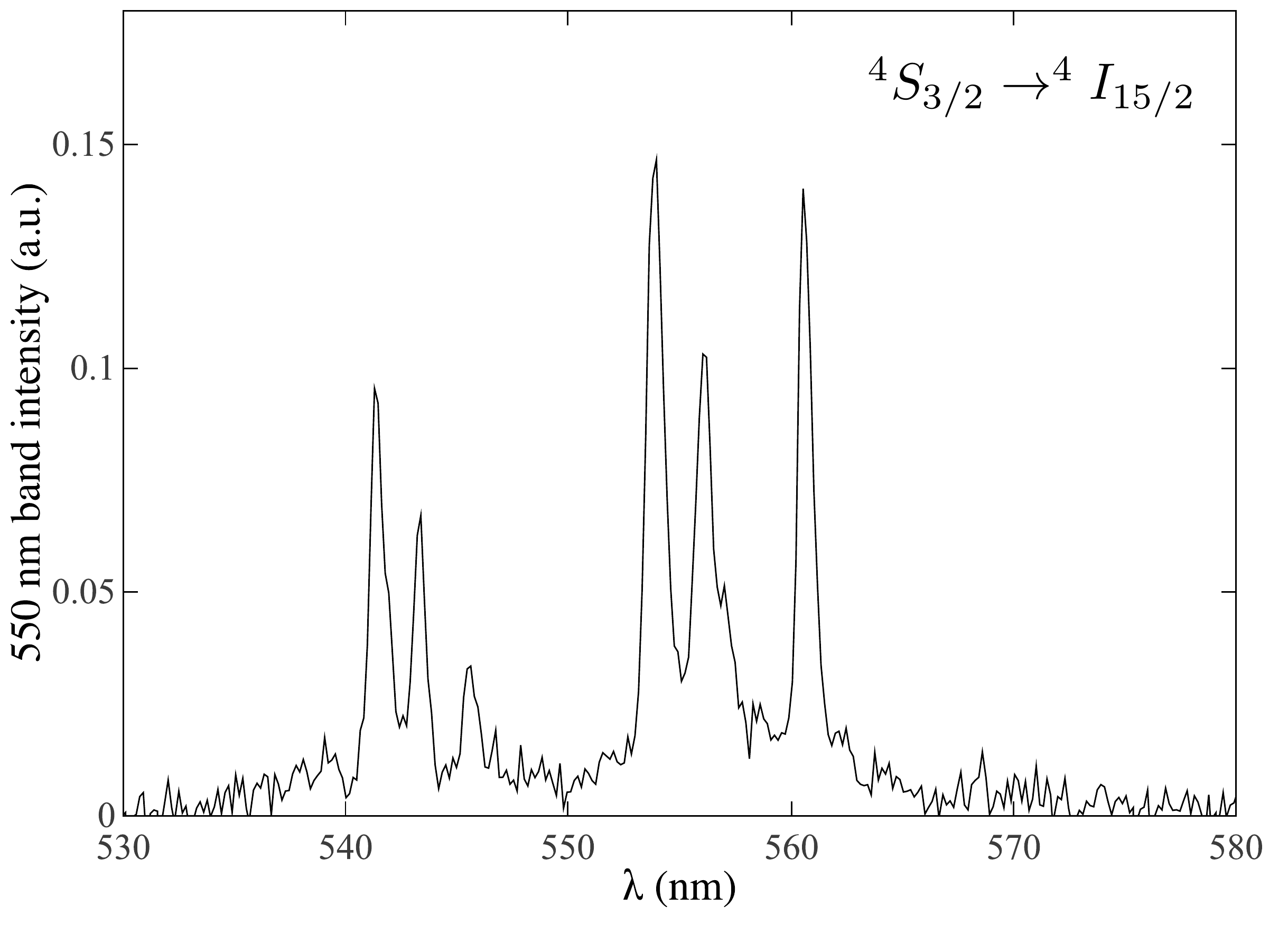}
\caption{
$540$\,nm fluorescence spectrum originating from the double resonance with a laser diode at wavelength 960\,nm and a Ti:Al$_2$O$_3$ laser.}
\label{540band}
\end{center}
\end{figure}

To study the IRQC efficiency, we have used as infrared source a 960\,nm wavelength diode laser.  Upconversion is accomplished by pumping the $^{4}I_{13/2} \rightarrow ^{4}S_{3/2}$ transition with a tunable Ti:Al$_2$O$_3$ laser. 
In Fig.\,\ref{540band} the fluorescence spectrum, obtained with a CCD spectrometer (Ocean Optics mod. Red ), around 540\,nm is reported.
We estimate that the IRQC efficiency is of the order of $10^{-4}$ with a pump flux of the order of 10\,W/cm$^{2}$. Such a low efficiency is related to the properties of the host material, as reported in Ref.\,\onlinecite{esterowitz:1968}.

\begin{figure}[h!]
\begin{center}
\includegraphics[width=2.8in]{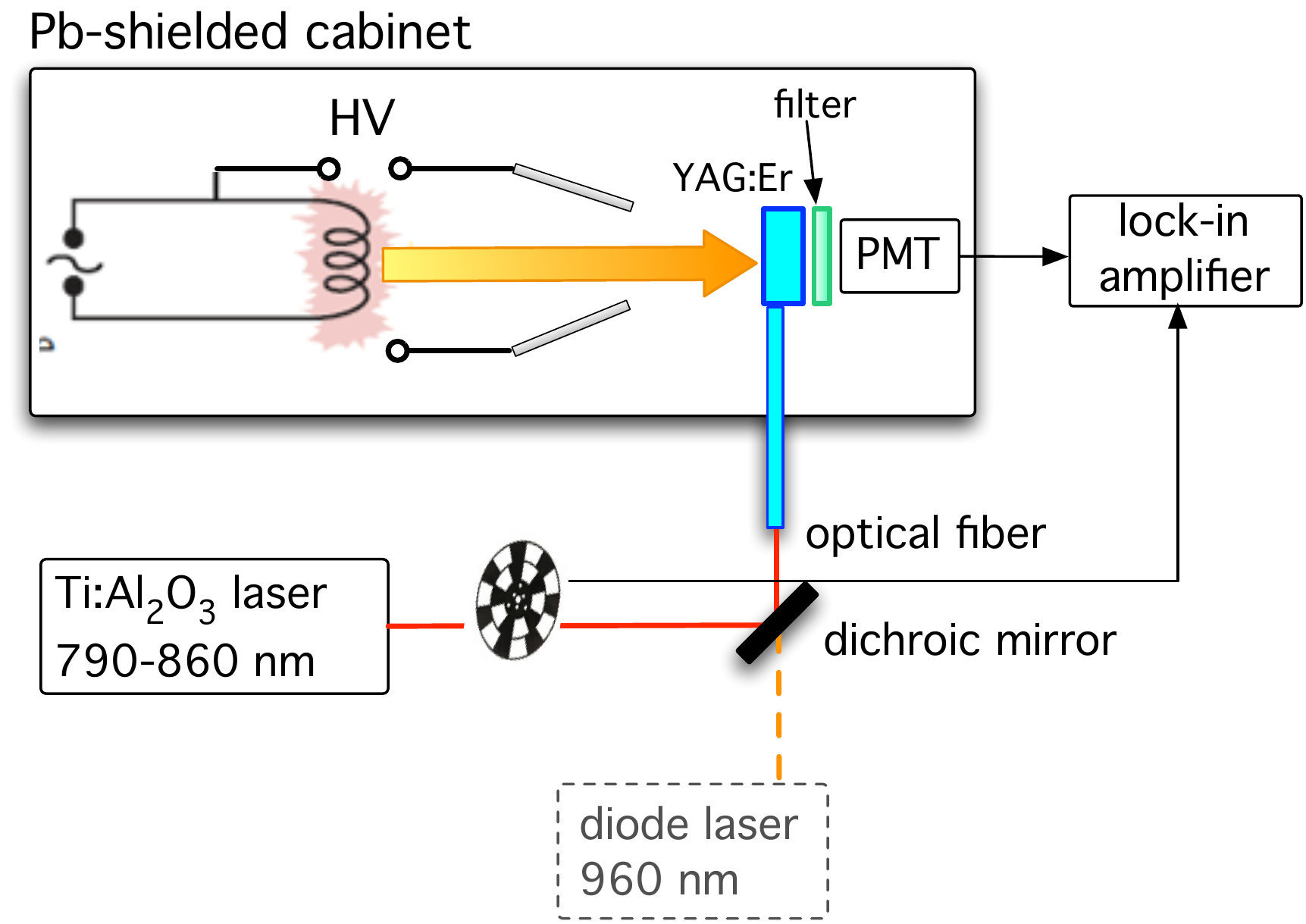}
\caption{
Scheme of the experimental apparatus for particle detection.
The output of a tunable Ti:Al$_2$O$_3$ laser is sent through a fiber to pump an YAG:Er crystal irradiated by $\sim60$ keV electrons. In the preliminary measurements, radiation at a wavelength of $\lambda=960\,$nm is coupled to the crystal together with the pump radiation in order to study the 540\,nm fluorescence band.
}
\label{sch}
\end{center}
\end{figure}

The experimental apparatus depicted in Fig.\,\ref{sch} was designed to test the particle detection through the IRQC scheme. A few hundred nA continuous current of 60 keV electrons impinges on the YAG:Er crystal while a Ti:Al$_2$O$_3$ laser pumped the transition $^{4}I_{13/2} \rightarrow ^{4}S_{3/2}$ as shown in the energy level scheme in Fig.\,\ref{3liv}. Pump light was allowed to impinge on the crystal by means of a 1.6\,mm diameter multimode fiber. 
 Detection of the fluorescence for the different laser pump wavelengths is accomplished by means of a lock-in amplifier connected to the output of a photomultiplier tube (PMT) and modulation of the intensity of the pump laser. 
 A bandpass filter allows the PMT to collect the total intensity emitted only in the range 540-560\,nm. 
 
\begin{figure}
\begin{center}
\includegraphics[width=3.2in]{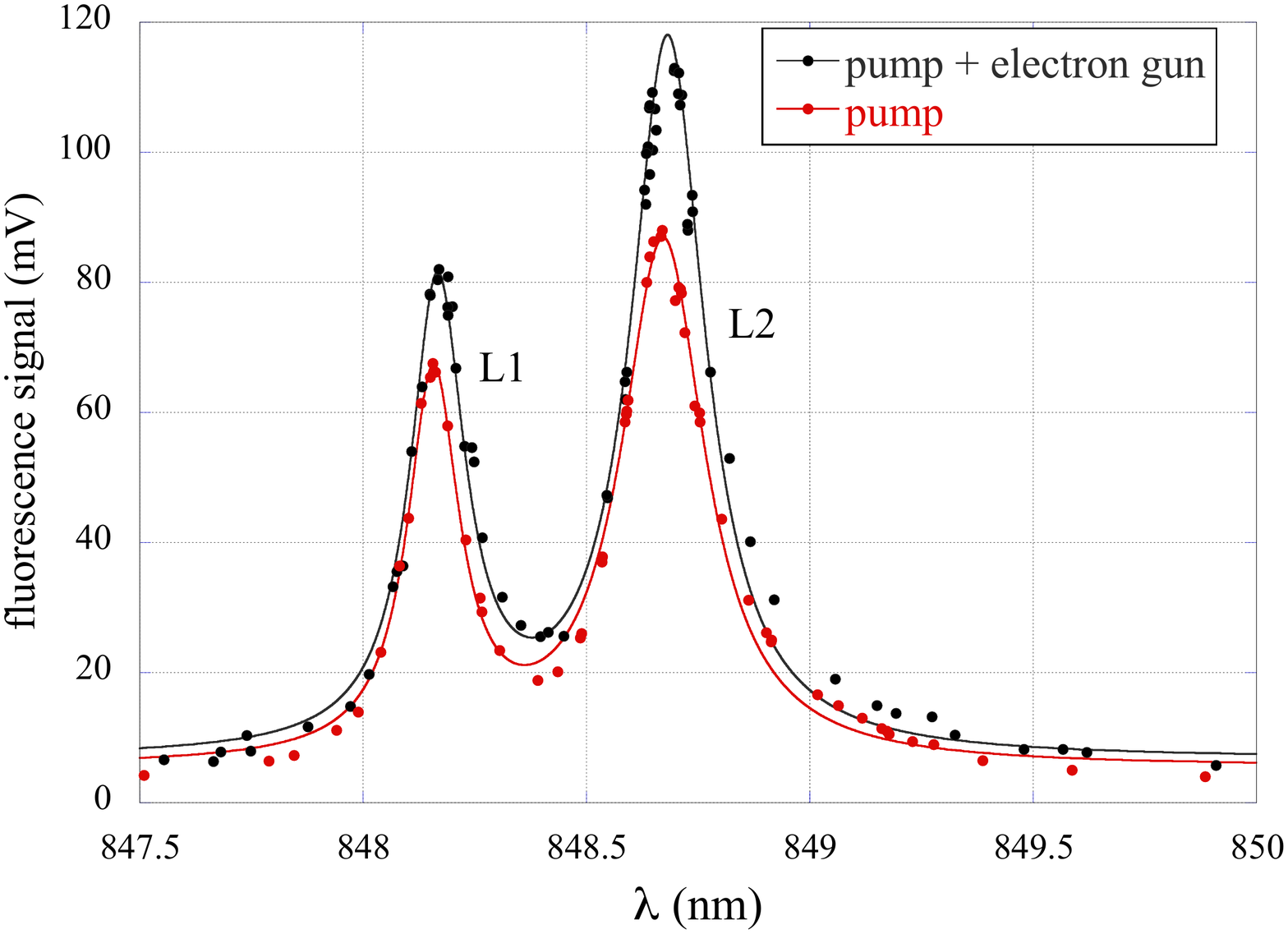}
\caption{Demonstration of the novel detector based on the IRQC concept: the 
540\,nm band fluorescence signal versus pump laser wavelength is greater when the electron gun excites the crystal. }
\label{two}
\end{center}
\end{figure}

\begin{figure}[h!]
\begin{center}
\includegraphics[width=1.8in]{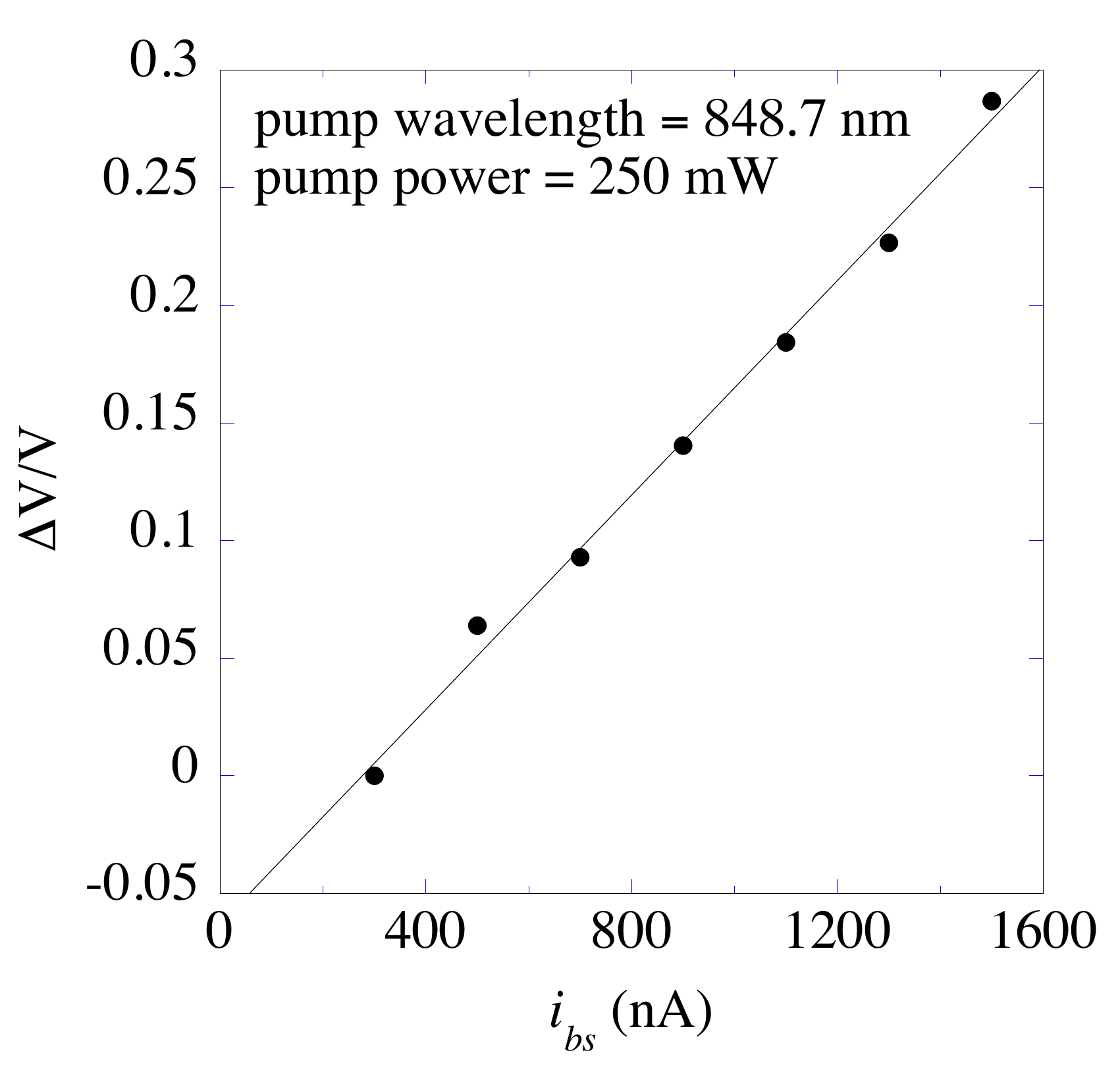}
\caption{The relative increase of the 540\,nm band intensity $\Delta V$ is linearly dependent on the current measured at the beam stopper.}
\label{lin1}
\end{center}
\end{figure}

\begin{figure}
\begin{center}
\includegraphics[width=3.0 in]{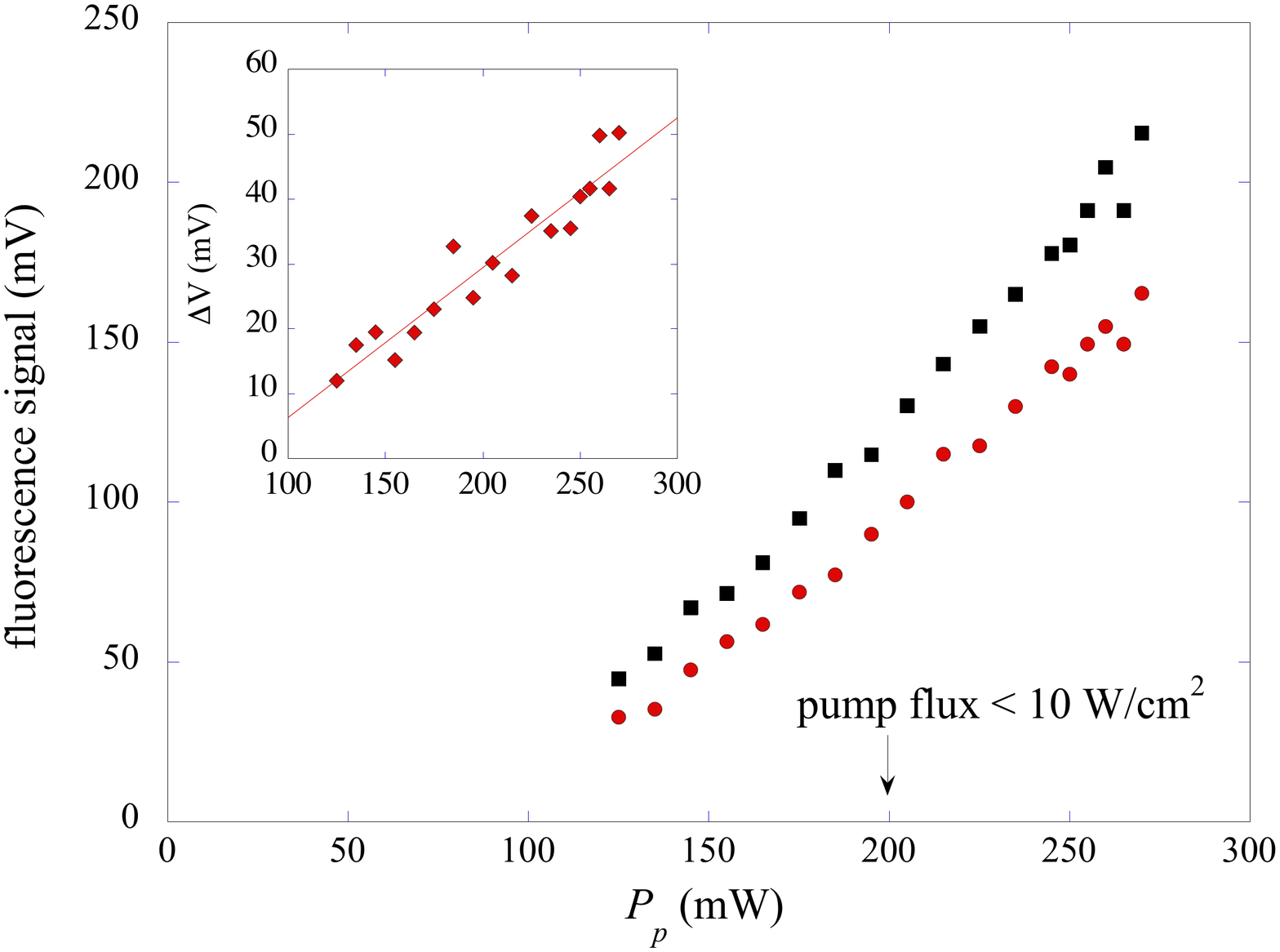}
\caption{Linear dependence of the fluorescence on the incident power. The filled dot symbol correspond to the data of the double resonance with the pump laser only, the filled squares represent the combination of the electron gun and the pump laser. }
\label{lin2}
\end{center}
\end{figure}

  In Fig.\,\ref{two} the $^{4}$S$_{3/2}$ fluorescence intensity is plotted for different wavelengths of the pump laser. 
    The measurement is repeated in the same conditions with the electron gun switched off in order to quantify the contribution of the double resonance with the pump laser only. The interaction of the electrons in the YAG:Er crystal can be discerned due to a 30\% increase of the areas under the Lorentzian curves. It is worth noticing that the contribution to the overall fluorescence due to the electrons excitation is geometrically unfavorite as compared to the laser double resonance because of the electron short range in the crystal and of the crystal absorption.
  Results of the double Lorentzian fit of the data are resumed in Table\,\ref{resume}.
   \begin{table}
\caption{\label{resume} Results of the Lorentizan fit. $w$ and $a$ are respectively the width of the peaks, and their area.}
\begin{ruledtabular}
\begin{tabular}{l|c|c|c|c}
  & & $\lambda_{max}$ & $w$ & $a$   \\
  \hline
  L1 & pump & $848.16\pm 0.01$ & $0.14\pm0.01$ & $12.8\pm0.8$ \\
 & pump + $e^-$gun & $848.17\pm 0.02$ & $0.15\pm0.01$ & $16.5\pm1.1$ \\ 
 \hline
  L2 & pump & $848.67\pm 0.01$ & $0.23\pm0.01$ &$28.8\pm1.1$ \\
 & pump + $e^-$gun & $848.68\pm 0.01$ & $0.2\pm0.01$ &$34.8\pm1.1$ \\  
\end{tabular}
\end{ruledtabular}
\end{table} 
As we apply lock-in techniques through a modulation of the pump signal, it is straightforward that no signal above the PMT threshold is detected when the electron gun is on.
The two lines correspond to two well defined transitions \cite{kaminski} between sublevels in the $^{4}$I$_{13/2}$ and the  $^{4}$S$_{3/2}$ manifolds. 
We have also checked that the fluorescence signal depends linearly on the electron gun current in a range $0.1-1.6\,\mu$A, as shown in Fig.\,\ref{lin1}. 
It is linear with the pump as well (Fig.\,\ref{lin2}), as observed up to the maximum available light intensity ($\sim 280$\,mW).

In conclusion, we have shown that a laser-pumped Er$^{3+}$-doped crystal emits a fluorescence signal in the 540\,nm band whose intensity is proportional to the energy deposited in the crystal. 
To the best of our knowledge, the results presented in this work represent the first demonstration of an IRQC scheme applied to the detection of particles. 
Such an encouraging result has been obtained in YAG:Er, a crystal that for our purposes cannot be considered as ideal. 
The optimum crystal is preferably transparent to the pump until a particle interacts in the activated material and deposits its energy in the volume shined by the pump laser. This is not the present case, as shown in Fig.\,\ref{two}, where a significant fraction of the fluorescence is determined by the double resonance with the pump laser.
Moreover, the host crystal (YAG) gives a much weaker IRQC output than in fluoride or tungstate hosts \cite{esterowitz:1968}. Another key requirement is the lifetime of the metastable level, and materials characterized by a much longer $\tau$ have been recently investigated, such as neodymium doped potassium lead bromide (KPb$_2$Br$_5$:Nd) in wich $\tau=57\,$ms at 15\,K \cite{Brown:2013fk}. This material presents another interesting aspect, because of its fluorescence at 1064\,nm that decays to the metastable level $^4I_{11/2}$, offering the possibility to exploit looping cycles to increase the detection sensitivity \cite{auzel:2004}.


\begin{thebibliography}{26}%
\makeatletter
\providecommand \@ifxundefined [1]{%
 \@ifx{#1\undefined}
}%
\providecommand \@ifnum [1]{%
 \ifnum #1\expandafter \@firstoftwo
 \else \expandafter \@secondoftwo
 \fi
}%
\providecommand \@ifx [1]{%
 \ifx #1\expandafter \@firstoftwo
 \else \expandafter \@secondoftwo
 \fi
}%
\providecommand \natexlab [1]{#1}%
\providecommand \enquote  [1]{``#1''}%
\providecommand \bibnamefont  [1]{#1}%
\providecommand \bibfnamefont [1]{#1}%
\providecommand \citenamefont [1]{#1}%
\providecommand \href@noop [0]{\@secondoftwo}%
\providecommand \href [0]{\begingroup \@sanitize@url \@href}%
\providecommand \@href[1]{\@@startlink{#1}\@@href}%
\providecommand \@@href[1]{\endgroup#1\@@endlink}%
\providecommand \@sanitize@url [0]{\catcode `\\12\catcode `\$12\catcode
  `\&12\catcode `\#12\catcode `\^12\catcode `\_12\catcode `\%12\relax}%
\providecommand \@@startlink[1]{}%
\providecommand \@@endlink[0]{}%
\providecommand \url  [0]{\begingroup\@sanitize@url \@url }%
\providecommand \@url [1]{\endgroup\@href {#1}{\urlprefix }}%
\providecommand \urlprefix  [0]{URL }%
\providecommand \Eprint [0]{\href }%
\providecommand \doibase [0]{http://dx.doi.org/}%
\providecommand \selectlanguage [0]{\@gobble}%
\providecommand \bibinfo  [0]{\@secondoftwo}%
\providecommand \bibfield  [0]{\@secondoftwo}%
\providecommand \translation [1]{[#1]}%
\providecommand \BibitemOpen [0]{}%
\providecommand \bibitemStop [0]{}%
\providecommand \bibitemNoStop [0]{.\EOS\space}%
\providecommand \EOS [0]{\spacefactor3000\relax}%
\providecommand \BibitemShut  [1]{\csname bibitem#1\endcsname}%
\let\auto@bib@innerbib\@empty
\bibitem [{\citenamefont {Sikivie}(2014)}]{Sikivie:2014fk}%
  \BibitemOpen
  \bibfield  {author} {\bibinfo {author} {\bibfnamefont {P.}~\bibnamefont
  {Sikivie}},\ }\href {http://link.aps.org/doi/10.1103/PhysRevLett.113.201301}
  {\bibfield  {journal} {\bibinfo  {journal} {Phys. Rev. Lett.}\ }\textbf
  {\bibinfo {volume} {113}},\ \bibinfo {pages} {201301--} (\bibinfo {year}
  {2014})}\BibitemShut {NoStop}%
\bibitem [{\citenamefont {Dodd}, \citenamefont {Papageorgiu},\ and\
  \citenamefont {Ranfone}(1991)}]{Dodd:1991uq}%
  \BibitemOpen
  \bibfield  {author} {\bibinfo {author} {\bibfnamefont {A.~C.}\ \bibnamefont
  {Dodd}}, \bibinfo {author} {\bibfnamefont {E.}~\bibnamefont {Papageorgiu}}, \
  and\ \bibinfo {author} {\bibfnamefont {S.}~\bibnamefont {Ranfone}},\ }\href
  {\doibase http://dx.doi.org/10.1016/0370-2693(91)91064-3} {\bibfield
  {journal} {\bibinfo  {journal} {Phys. Lett. B}\ }\textbf {\bibinfo {volume}
  {266}},\ \bibinfo {pages} {434--438} (\bibinfo {year} {1991})}\BibitemShut
  {NoStop}%
\bibitem [{\citenamefont {Freedman}(1974)}]{Freedman:1974fk}%
  \BibitemOpen
  \bibfield  {author} {\bibinfo {author} {\bibfnamefont {D.~Z.}\ \bibnamefont
  {Freedman}},\ }\href {http://link.aps.org/doi/10.1103/PhysRevD.9.1389}
  {\bibfield  {journal} {\bibinfo  {journal} {Phys. Rev. D}\ }\textbf {\bibinfo
  {volume} {9}},\ \bibinfo {pages} {1389--1392} (\bibinfo {year}
  {1974})}\BibitemShut {NoStop}%
\bibitem [{\citenamefont {Aalseth}\ \emph {et~al.}(2011)\citenamefont
  {Aalseth}, \citenamefont {Barbeau}, \citenamefont {Bowden}, \citenamefont
  {Cabrera-Palmer}, \citenamefont {Colaresi}, \citenamefont {Collar},
  \citenamefont {Dazeley}, \citenamefont {de~Lurgio}, \citenamefont {Fast},
  \citenamefont {Fields}, \citenamefont {Greenberg}, \citenamefont {Hossbach},
  \citenamefont {Keillor}, \citenamefont {Kephart}, \citenamefont {Marino},
  \citenamefont {Miley}, \citenamefont {Miller}, \citenamefont {Orrell},
  \citenamefont {Radford}, \citenamefont {Reyna}, \citenamefont {Tench},
  \citenamefont {Van~Wechel}, \citenamefont {Wilkerson},\ and\ \citenamefont
  {Yocum}}]{Aalseth:2011fk}%
  \BibitemOpen
  \bibfield  {author} {\bibinfo {author} {\bibfnamefont {C.~E.}\ \bibnamefont
  {Aalseth}}, \bibinfo {author} {\bibfnamefont {P.~S.}\ \bibnamefont
  {Barbeau}}, \bibinfo {author} {\bibfnamefont {N.~S.}\ \bibnamefont {Bowden}},
  \bibinfo {author} {\bibfnamefont {B.}~\bibnamefont {Cabrera-Palmer}},
  \bibinfo {author} {\bibfnamefont {J.}~\bibnamefont {Colaresi}}, \bibinfo
  {author} {\bibfnamefont {J.~I.}\ \bibnamefont {Collar}}, \bibinfo {author}
  {\bibfnamefont {S.}~\bibnamefont {Dazeley}}, \bibinfo {author} {\bibfnamefont
  {P.}~\bibnamefont {de~Lurgio}}, \bibinfo {author} {\bibfnamefont {J.~E.}\
  \bibnamefont {Fast}}, \bibinfo {author} {\bibfnamefont {N.}~\bibnamefont
  {Fields}}, \bibinfo {author} {\bibfnamefont {C.~H.}\ \bibnamefont
  {Greenberg}}, \bibinfo {author} {\bibfnamefont {T.~W.}\ \bibnamefont
  {Hossbach}}, \bibinfo {author} {\bibfnamefont {M.~E.}\ \bibnamefont
  {Keillor}}, \bibinfo {author} {\bibfnamefont {J.~D.}\ \bibnamefont
  {Kephart}}, \bibinfo {author} {\bibfnamefont {M.~G.}\ \bibnamefont {Marino}},
  \bibinfo {author} {\bibfnamefont {H.~S.}\ \bibnamefont {Miley}}, \bibinfo
  {author} {\bibfnamefont {M.~L.}\ \bibnamefont {Miller}}, \bibinfo {author}
  {\bibfnamefont {J.~L.}\ \bibnamefont {Orrell}}, \bibinfo {author}
  {\bibfnamefont {D.~C.}\ \bibnamefont {Radford}}, \bibinfo {author}
  {\bibfnamefont {D.}~\bibnamefont {Reyna}}, \bibinfo {author} {\bibfnamefont
  {O.}~\bibnamefont {Tench}}, \bibinfo {author} {\bibfnamefont {T.~D.}\
  \bibnamefont {Van~Wechel}}, \bibinfo {author} {\bibfnamefont {J.~F.}\
  \bibnamefont {Wilkerson}}, \ and\ \bibinfo {author} {\bibfnamefont {K.~M.}\
  \bibnamefont {Yocum}},\ }\href
  {http://link.aps.org/doi/10.1103/PhysRevLett.106.131301} {\bibfield
  {journal} {\bibinfo  {journal} {Phys. Rev. Lett.}\ }\textbf {\bibinfo
  {volume} {106}},\ \bibinfo {pages} {131301--} (\bibinfo {year}
  {2011})}\BibitemShut {NoStop}%
\bibitem [{\citenamefont {Enss}(2005)}]{enss:2005}%
  \BibitemOpen
  \bibfield  {author} {\bibinfo {author} {\bibfnamefont {C.}~\bibnamefont
  {Enss}},\ }\href@noop {} {\emph {\bibinfo {title} {Cryogenic Particle
  Detection}}}\ (\bibinfo  {publisher} {Springer-Verlag Berlin Heidelberg},\
  \bibinfo {year} {2005})\BibitemShut {NoStop}%
\bibitem [{\citenamefont {Bloembergen}(1959)}]{blo:1959}%
  \BibitemOpen
  \bibfield  {author} {\bibinfo {author} {\bibfnamefont {N.}~\bibnamefont
  {Bloembergen}},\ }\href@noop {} {\bibfield  {journal} {\bibinfo  {journal}
  {Phys. Rev. Lett}\ }\textbf {\bibinfo {volume} {2}},\ \bibinfo {pages} {84}
  (\bibinfo {year} {1959})}\BibitemShut {NoStop}%
\bibitem [{\citenamefont {Kaminskii}(1981)}]{kaminski}%
  \BibitemOpen
  \bibfield  {author} {\bibinfo {author} {\bibfnamefont {A.~K.}\ \bibnamefont
  {Kaminskii}},\ }\href@noop {} {\emph {\bibinfo {title} {Laser crystals}}}\
  (\bibinfo  {publisher} {Springer-Verlag Berlin},\ \bibinfo {year}
  {1981})\BibitemShut {NoStop}%
\bibitem [{\citenamefont {Gruber}\ \emph {et~al.}(1993)\citenamefont {Gruber},
  \citenamefont {Quagliano}, \citenamefont {Reid}, \citenamefont {Richardson},
  \citenamefont {Hills}, \citenamefont {Seltzer}, \citenamefont {Stevens},
  \citenamefont {Morrison},\ and\ \citenamefont {Allik}}]{gruber:1993}%
  \BibitemOpen
  \bibfield  {author} {\bibinfo {author} {\bibfnamefont {J.~B.}\ \bibnamefont
  {Gruber}}, \bibinfo {author} {\bibfnamefont {J.~R.}\ \bibnamefont
  {Quagliano}}, \bibinfo {author} {\bibfnamefont {M.~F.}\ \bibnamefont {Reid}},
  \bibinfo {author} {\bibfnamefont {F.~S.}\ \bibnamefont {Richardson}},
  \bibinfo {author} {\bibfnamefont {M.~E.}\ \bibnamefont {Hills}}, \bibinfo
  {author} {\bibfnamefont {M.~D.}\ \bibnamefont {Seltzer}}, \bibinfo {author}
  {\bibfnamefont {S.~B.}\ \bibnamefont {Stevens}}, \bibinfo {author}
  {\bibfnamefont {C.~A.}\ \bibnamefont {Morrison}}, \ and\ \bibinfo {author}
  {\bibfnamefont {T.~H.}\ \bibnamefont {Allik}},\ }\href@noop {} {\bibfield
  {journal} {\bibinfo  {journal} {Phys. Rev. B}\ }\textbf {\bibinfo {volume}
  {48}},\ \bibinfo {pages} {15561--15573} (\bibinfo {year} {1993})}\BibitemShut
  {NoStop}%
\bibitem [{\citenamefont {Rodnyi}(1997)}]{Rodnyi}%
  \BibitemOpen
  \bibfield  {author} {\bibinfo {author} {\bibfnamefont {P.~A.}\ \bibnamefont
  {Rodnyi}},\ }\href@noop {} {}\ (\bibinfo  {publisher} {CRC Press LLC},\
  \bibinfo {year} {1997})\BibitemShut {NoStop}%
\bibitem [{\citenamefont {Weber}(2002)}]{Weber200235}%
  \BibitemOpen
  \bibfield  {author} {\bibinfo {author} {\bibfnamefont {M.~J.}\ \bibnamefont
  {Weber}},\ }\href {\doibase http://dx.doi.org/10.1016/S0022-2313(02)00423-4}
  {\bibfield  {journal} {\bibinfo  {journal} {J. Lumin.}\ }\textbf {\bibinfo
  {volume} {100}},\ \bibinfo {pages} {35 -- 45} (\bibinfo {year}
  {2002})}\BibitemShut {NoStop}%
\bibitem [{\citenamefont {Drozdowski}\ \emph {et~al.}(2008)\citenamefont
  {Drozdowski}, \citenamefont {Dorenbos}, \citenamefont {de~Haas},
  \citenamefont {Drozdowska}, \citenamefont {Owens}, \citenamefont {Kamada},
  \citenamefont {Tsutsumi}, \citenamefont {Usuki}, \citenamefont {Yanagida},\
  and\ \citenamefont {Yoshikawa}}]{Drozdowski:2008}%
  \BibitemOpen
  \bibfield  {author} {\bibinfo {author} {\bibfnamefont {W.}~\bibnamefont
  {Drozdowski}}, \bibinfo {author} {\bibfnamefont {P.}~\bibnamefont
  {Dorenbos}}, \bibinfo {author} {\bibfnamefont {J.}~\bibnamefont {de~Haas}},
  \bibinfo {author} {\bibfnamefont {R.}~\bibnamefont {Drozdowska}}, \bibinfo
  {author} {\bibfnamefont {A.}~\bibnamefont {Owens}}, \bibinfo {author}
  {\bibfnamefont {K.}~\bibnamefont {Kamada}}, \bibinfo {author} {\bibfnamefont
  {K.}~\bibnamefont {Tsutsumi}}, \bibinfo {author} {\bibfnamefont
  {Y.}~\bibnamefont {Usuki}}, \bibinfo {author} {\bibfnamefont
  {T.}~\bibnamefont {Yanagida}}, \ and\ \bibinfo {author} {\bibfnamefont
  {A.}~\bibnamefont {Yoshikawa}},\ }\href@noop {} {\bibfield  {journal}
  {\bibinfo  {journal} {IEEE Trans. Nucl. Sci.}\ }\textbf {\bibinfo {volume}
  {55}},\ \bibinfo {pages} {2420--2424} (\bibinfo {year} {2008})}\BibitemShut
  {NoStop}%
\bibitem [{\citenamefont {Moses}\ \emph {et~al.}(1998)\citenamefont {Moses},
  \citenamefont {Weber}, \citenamefont {Derenzo}, \citenamefont {Perry},
  \citenamefont {Berdahl},\ and\ \citenamefont {Boatner}}]{moses:1998}%
  \BibitemOpen
  \bibfield  {author} {\bibinfo {author} {\bibfnamefont {W.}~\bibnamefont
  {Moses}}, \bibinfo {author} {\bibfnamefont {M.}~\bibnamefont {Weber}},
  \bibinfo {author} {\bibfnamefont {S.}~\bibnamefont {Derenzo}}, \bibinfo
  {author} {\bibfnamefont {D.}~\bibnamefont {Perry}}, \bibinfo {author}
  {\bibfnamefont {P.}~\bibnamefont {Berdahl}}, \ and\ \bibinfo {author}
  {\bibfnamefont {L.}~\bibnamefont {Boatner}},\ }\href@noop {} {\bibfield
  {journal} {\bibinfo  {journal} {IEEE Trans. Nucl. Sci.}\ }\textbf {\bibinfo
  {volume} {45}} (\bibinfo {year} {1998})}\BibitemShut {NoStop}%
\bibitem [{\citenamefont {Knoll}(2010)}]{knoll}%
  \BibitemOpen
  \bibfield  {author} {\bibinfo {author} {\bibfnamefont {G.~F.}\ \bibnamefont
  {Knoll}},\ }\href@noop {} {\emph {\bibinfo {title} {Radiation detection and
  measurement}}}\ (\bibinfo  {publisher} {Wiley},\ \bibinfo {address} {New
  York},\ \bibinfo {year} {2010})\BibitemShut {NoStop}%
\bibitem [{\citenamefont {Antonini}\ \emph {et~al.}(2002)\citenamefont
  {Antonini}, \citenamefont {Belogurov}, \citenamefont {Bressi}, \citenamefont
  {Carugno},\ and\ \citenamefont {Iannuzzi}}]{Antonini:2002fk}%
  \BibitemOpen
  \bibfield  {author} {\bibinfo {author} {\bibfnamefont {P.}~\bibnamefont
  {Antonini}}, \bibinfo {author} {\bibfnamefont {S.}~\bibnamefont {Belogurov}},
  \bibinfo {author} {\bibfnamefont {G.}~\bibnamefont {Bressi}}, \bibinfo
  {author} {\bibfnamefont {G.}~\bibnamefont {Carugno}}, \ and\ \bibinfo
  {author} {\bibfnamefont {D.}~\bibnamefont {Iannuzzi}},\ }\href {\doibase
  http://dx.doi.org/10.1016/S0168-9002(01)02164-7} {\bibfield  {journal}
  {\bibinfo  {journal} {Nucl. Instrum. Meth. A}\ }\textbf {\bibinfo {volume}
  {486}},\ \bibinfo {pages} {799--802} (\bibinfo {year} {2002})}\BibitemShut
  {NoStop}%
\bibitem [{\citenamefont {Krupke}(1965)}]{krupke:1965}%
  \BibitemOpen
  \bibfield  {author} {\bibinfo {author} {\bibfnamefont {W.~F.}\ \bibnamefont
  {Krupke}},\ }\href@noop {} {\bibfield  {journal} {\bibinfo  {journal} {IEEE
  J. Quant. Electron.}\ ,\ \bibinfo {pages} {20}} (\bibinfo {year}
  {1965})}\BibitemShut {NoStop}%
\bibitem [{\citenamefont {Wright}\ \emph {et~al.}(1973)\citenamefont {Wright},
  \citenamefont {Zalucha}, \citenamefont {Lauer}, \citenamefont {Cox},\ and\
  \citenamefont {Fong}}]{wright:1973}%
  \BibitemOpen
  \bibfield  {author} {\bibinfo {author} {\bibfnamefont {J.~C.}\ \bibnamefont
  {Wright}}, \bibinfo {author} {\bibfnamefont {D.~J.}\ \bibnamefont {Zalucha}},
  \bibinfo {author} {\bibfnamefont {H.~V.}\ \bibnamefont {Lauer}}, \bibinfo
  {author} {\bibfnamefont {D.~E.}\ \bibnamefont {Cox}}, \ and\ \bibinfo
  {author} {\bibfnamefont {F.~K.}\ \bibnamefont {Fong}},\ }\href@noop {}
  {\bibfield  {journal} {\bibinfo  {journal} {J. Appl. Phys.}\ }\textbf
  {\bibinfo {volume} {44}},\ \bibinfo {pages} {781} (\bibinfo {year}
  {1973})}\BibitemShut {NoStop}%
\bibitem [{\citenamefont {Barcellan}\ \emph {et~al.}(2011)\citenamefont
  {Barcellan}, \citenamefont {Berto}, \citenamefont {Carugno}, \citenamefont
  {Galet}, \citenamefont {Galeazzi},\ and\ \citenamefont
  {Borghesani}}]{borghe:2011}%
  \BibitemOpen
  \bibfield  {author} {\bibinfo {author} {\bibfnamefont {L.}~\bibnamefont
  {Barcellan}}, \bibinfo {author} {\bibfnamefont {E.}~\bibnamefont {Berto}},
  \bibinfo {author} {\bibfnamefont {G.}~\bibnamefont {Carugno}}, \bibinfo
  {author} {\bibfnamefont {G.}~\bibnamefont {Galet}}, \bibinfo {author}
  {\bibfnamefont {G.}~\bibnamefont {Galeazzi}}, \ and\ \bibinfo {author}
  {\bibfnamefont {A.~F.}\ \bibnamefont {Borghesani}},\ }\href@noop {}
  {\bibfield  {journal} {\bibinfo  {journal} {Rev. Sci. Instrum.}\ }\textbf
  {\bibinfo {volume} {82}},\ \bibinfo {pages} {95103} (\bibinfo {year}
  {2011})}\BibitemShut {NoStop}%
\bibitem [{\citenamefont {Zorenko}\ \emph {et~al.}(2014)\citenamefont
  {Zorenko}, \citenamefont {Gorbenko}, \citenamefont {Zorenko}, \citenamefont
  {Savchyn}, \citenamefont {Batentschuk}, \citenamefont {Osvet},\ and\
  \citenamefont {Brabec}}]{Zorenko:2014fk}%
  \BibitemOpen
  \bibfield  {author} {\bibinfo {author} {\bibfnamefont {Y.}~\bibnamefont
  {Zorenko}}, \bibinfo {author} {\bibfnamefont {V.}~\bibnamefont {Gorbenko}},
  \bibinfo {author} {\bibfnamefont {T.}~\bibnamefont {Zorenko}}, \bibinfo
  {author} {\bibfnamefont {V.}~\bibnamefont {Savchyn}}, \bibinfo {author}
  {\bibfnamefont {M.}~\bibnamefont {Batentschuk}}, \bibinfo {author}
  {\bibfnamefont {A.}~\bibnamefont {Osvet}}, \ and\ \bibinfo {author}
  {\bibfnamefont {C.}~\bibnamefont {Brabec}},\ }\href {\doibase
  http://dx.doi.org/10.1016/j.jlumin.2014.04.025} {\bibfield  {journal}
  {\bibinfo  {journal} {J. Lumin.}\ }\textbf {\bibinfo {volume} {154}},\
  \bibinfo {pages} {198--203} (\bibinfo {year} {2014})}\BibitemShut {NoStop}%
\bibitem [{nis()}]{nist}%
  \BibitemOpen
  \href {{physics.nist.gov/PhysRevData/Star/Text/method.html}} {\enquote
  {\bibinfo {title} {{physics.nist.gov/PhysRevData/Star/Text/method.html}},}\
  }\BibitemShut {NoStop}%
\bibitem [{\citenamefont {Eichhorn}\ \emph {et~al.}(2008)\citenamefont
  {Eichhorn}, \citenamefont {Fredrich-Thornton}, \citenamefont {Heumann},\ and\
  \citenamefont {Huber}}]{eich:2008}%
  \BibitemOpen
  \bibfield  {author} {\bibinfo {author} {\bibfnamefont {M.}~\bibnamefont
  {Eichhorn}}, \bibinfo {author} {\bibfnamefont {S.~T.}\ \bibnamefont
  {Fredrich-Thornton}}, \bibinfo {author} {\bibfnamefont {E.}~\bibnamefont
  {Heumann}}, \ and\ \bibinfo {author} {\bibfnamefont {G.}~\bibnamefont
  {Huber}},\ }\href@noop {} {\bibfield  {journal} {\bibinfo  {journal} {Appl.
  Phys. B}\ }\textbf {\bibinfo {volume} {91}},\ \bibinfo {pages} {249}
  (\bibinfo {year} {2008})}\BibitemShut {NoStop}%
\bibitem [{\citenamefont {Payne}\ \emph {et~al.}(1992)\citenamefont {Payne},
  \citenamefont {Chase}, \citenamefont {Smith}, \citenamefont {Kway},\ and\
  \citenamefont {Krupke}}]{Payne:1992}%
  \BibitemOpen
  \bibfield  {author} {\bibinfo {author} {\bibfnamefont {S.}~\bibnamefont
  {Payne}}, \bibinfo {author} {\bibfnamefont {L.}~\bibnamefont {Chase}},
  \bibinfo {author} {\bibfnamefont {L.~K.}\ \bibnamefont {Smith}}, \bibinfo
  {author} {\bibfnamefont {W.~L.}\ \bibnamefont {Kway}}, \ and\ \bibinfo
  {author} {\bibfnamefont {W.~F.}\ \bibnamefont {Krupke}},\ }\href@noop {}
  {\bibfield  {journal} {\bibinfo  {journal} {IEEE J. Quant. Electron.}\
  }\textbf {\bibinfo {volume} {28}},\ \bibinfo {pages} {2619--2630} (\bibinfo
  {year} {1992})}\BibitemShut {NoStop}%
\bibitem [{\citenamefont {Ter-Gabrielyan}\ \emph {et~al.}(2009)\citenamefont
  {Ter-Gabrielyan}, \citenamefont {Dubinskii}, \citenamefont {Newburgh},
  \citenamefont {Michael},\ and\ \citenamefont {Merkle}}]{Ter-Gabrielyan:2009}%
  \BibitemOpen
  \bibfield  {author} {\bibinfo {author} {\bibfnamefont {N.}~\bibnamefont
  {Ter-Gabrielyan}}, \bibinfo {author} {\bibfnamefont {M.}~\bibnamefont
  {Dubinskii}}, \bibinfo {author} {\bibfnamefont {G.~A.}\ \bibnamefont
  {Newburgh}}, \bibinfo {author} {\bibfnamefont {A.}~\bibnamefont {Michael}}, \
  and\ \bibinfo {author} {\bibfnamefont {L.~D.}\ \bibnamefont {Merkle}},\
  }\href@noop {} {\bibfield  {journal} {\bibinfo  {journal} {Opt. Express}\
  }\textbf {\bibinfo {volume} {17}},\ \bibinfo {pages} {7159} (\bibinfo {year}
  {2009})}\BibitemShut {NoStop}%
\bibitem [{\citenamefont {Toma}\ and\ \citenamefont
  {Georgescu}(2006)}]{toma:2006}%
  \BibitemOpen
  \bibfield  {author} {\bibinfo {author} {\bibfnamefont {O.}~\bibnamefont
  {Toma}}\ and\ \bibinfo {author} {\bibfnamefont {S.}~\bibnamefont
  {Georgescu}},\ }\href@noop {} {\bibfield  {journal} {\bibinfo  {journal}
  {IEEE J. Quant. Electron.}\ }\textbf {\bibinfo {volume} {42}},\ \bibinfo
  {pages} {192} (\bibinfo {year} {2006})}\BibitemShut {NoStop}%
\bibitem [{\citenamefont {Esterowitz}, \citenamefont {A.~Schnitzler},\ and\
  \citenamefont {Bahler}(1968)}]{esterowitz:1968}%
  \BibitemOpen
  \bibfield  {author} {\bibinfo {author} {\bibfnamefont {L.}~\bibnamefont
  {Esterowitz}}, \bibinfo {author} {\bibfnamefont {J.~N.}\ \bibnamefont
  {A.~Schnitzler}}, \ and\ \bibinfo {author} {\bibfnamefont {J.}~\bibnamefont
  {Bahler}},\ }\href@noop {} {\bibfield  {journal} {\bibinfo  {journal} {Appl.
  Opt.}\ }\textbf {\bibinfo {volume} {7}},\ \bibinfo {pages} {2053} (\bibinfo
  {year} {1968})}\BibitemShut {NoStop}%
\bibitem [{\citenamefont {Brown}\ \emph {et~al.}(2013)\citenamefont {Brown},
  \citenamefont {Hanley}, \citenamefont {H{\"o}mmerich}, \citenamefont
  {Bluiett},\ and\ \citenamefont {Trivedi}}]{Brown:2013fk}%
  \BibitemOpen
  \bibfield  {author} {\bibinfo {author} {\bibfnamefont {E.}~\bibnamefont
  {Brown}}, \bibinfo {author} {\bibfnamefont {C.~B.}\ \bibnamefont {Hanley}},
  \bibinfo {author} {\bibfnamefont {U.}~\bibnamefont {H{\"o}mmerich}}, \bibinfo
  {author} {\bibfnamefont {A.~G.}\ \bibnamefont {Bluiett}}, \ and\ \bibinfo
  {author} {\bibfnamefont {S.~B.}\ \bibnamefont {Trivedi}},\ }\href {\doibase
  http://dx.doi.org/10.1016/j.jlumin.2011.12.023} {\bibfield  {journal}
  {\bibinfo  {journal} {J. Lumin.}\ }\textbf {\bibinfo {volume} {133}},\
  \bibinfo {pages} {244--248} (\bibinfo {year} {2013})}\BibitemShut {NoStop}%
\bibitem [{\citenamefont {Auzel}(2004)}]{auzel:2004}%
  \BibitemOpen
  \bibfield  {author} {\bibinfo {author} {\bibfnamefont {F.}~\bibnamefont
  {Auzel}},\ }\href@noop {} {\bibfield  {journal} {\bibinfo  {journal} {Chem.
  Rev.}\ }\textbf {\bibinfo {volume} {104}},\ \bibinfo {pages} {139--173}
  (\bibinfo {year} {2004})}\BibitemShut {NoStop}%
\end{thebibliography}


%

\end{document}